# Bias field correction of MPRAGE by an external reference – The poor man's MP2RAGE

Hampus Olsson[1], Gunther Helms[1]

*[1]Department. of Medical Radiation Physics, Clinical Sciences Lund, Lund University, Lund, Sweden.*

**Word count: 4687**

**Correpondence to:**

Hampus Olsson, M.Sc.

Medical Radiation Physics

Barngatan 4

Lund, SE 221 85

Sweden

+46737097806

Hampus.Olsson@med.lu.se

## Abstract

Purpose: To implement and evaluate a sequential approach to obtain semi-quantitative $T_1$-weighted MPRAGE images, unbiased by $B_1$ inhomogeneities at 7T.

Methods: In the reference gradient echo used for normalization of the MPRAGE image, flip angle ($\alpha_{GE}$) and acquisition voxel size ($V_{ref}$) was varied to optimize tissue contrast and acquisition time ($T_{acq}$). The finalized protocol was implemented at three different resolutions and the reproducibility was evaluated. Maps of $T_1$ were derived based on the normalized MPRAGE through forward signal modelling.

Results: A good compromise between tissue contrast and SNR was reached at $\alpha_{GE}$=3°. A reduction of the reference GE $T_{acq}$ by a factor of 4, at the cost of negligible bias, was obtained by increasing $V_{ref}$ with a factor of 8 relative the MPRAGE resolution. The coefficient-of-variation in segmented WM was 9±5% after normalization, compared to 24±12% before. The $T_1$ maps showed no obvious bias and had reasonable values with regard to literature, especially after optional $B_1$ correction through separate flip angle mapping.

Conclusion: A non-interleaved acquisition for normalization of MPRAGE offers a simple alternative to MP2RAGE to obtain semi-quantitative purely $T_1$-weighted images. These images can be converted to $T_1$ maps analogously to the established MP2RAGE approach. Scan time can be reduced by increasing $V_{ref}$ which has a miniscule effect on image quality.

## 1. Introduction

The magnetization prepared rapid gradient echo (MPRAGE) sequence has become the standard for structural $T_1$-weighted ($T_1$-w) 3D imaging. $T_1$ contrast is obtained through an inversion pulse followed by a rapid gradient echo (RAGE) readout with optional delays for free recovery before and after the readout (1). At ultra-high field (UHF) MRI of 7T or above, the SNR is improved through the increased polarization of nuclear spins, which can be translated into increased spatial resolution or faster scan times. The latter allows for visualization of substructures, unfeasible at lower field strengths (2). A major drawback of UHF is the increased inhomogeneity of the radiofrequency $B_1$ field. This applies to both the transmit ($B_1^+$) and receive sensitivity components, which combines to form a total bias field, corrupting the MR image. To correct for this intensity field bias, it has been suggested to acquire a reference gradient echo (GE) in conjunction with the MPRAGE sequence (3). Through a simple division of the two acquisitions, a normalized MPRAGE image is obtained, where signal variations due to the shared receive coils are eliminated. On the other hand, $B_1^+$ affects MPRAGE and GE differently through the local flip angle, hence the related bias is only reduced (not removed) from the normalized MPRAGE. The division also removes influence of proton density (PD) and $T_2$*, thus creating a purely $T_1$-w image with improved tissue contrast. The reference GE can be acquired either separately or interleaved with the MPRAGE at a longer inversion time (TI) where the latter approach has been popularized as the MP2RAGE sequence (4). The $B_1$ field will vary based on subject, positioning and coil (5). This means that pixel values from images acquired at different scanning sessions will normally not be comparable which is required for longitudinal or multi-site studies. Since it is independent of receive sensitivity and influence of $B_1^+$ inhomogeneity is reduced, this issue is addressed by the normalized MPRAGE image. The pixel values still lack physical meaning; hence the technique is considered “semi-quantitative”.

In this work, we describe the implementation of the non-interleaved variant at 7T, referred to as the “poor man’s MP2RAGE”. Although this variant is more susceptible to inter-scan subject motion, the reference GE can be accelerated through enlarged acquisition voxels, allowing for shorter scan time. The optimization procedure focused on the reference GE, specifically on the flip angle (3.1) and voxel size (3.2) in an effort to improve contrast, SNR, residual $B_1^+$ bias and scan time. The protocol was implemented at three different resolutions.

Normalized MPRAGE images acquired at a higher vs a lower resolution was compared (Methods 3.3). Further, the reproducibility (crucial for a semi-quantitative protocol) was also investigated (Methods 3.4). Lastly, the possibility to calculate $T_1$ maps through forward modelling to obtain a look-up table (LUT) of the normalized signal was explored (Methods 3.5). Thus, leaving the semi-quantitative domain to obtain fully quantitative maps which allows for a more direct biophysical interpretation in terms of, for instance, myelination (6). This is analogous to the approach described for the interleaved MP2RAGE by Marques et al. (4). This work ensured than an optimized protocol for bias field-corrected structural imaging, easily interpretable by radiologists, was supplied to the research site.

## 2. Theory

Influence of receive sensitivity, PD and $T_2$* was removed through division of the MPRAGE signal, $S_{MP}$, by the reference GE signal, $S_{GE}$, to yield the normalized MPRAGE signal, $S_{MP/GE}$. The rationale is illustrated by paraphrasing Eq. (3) in ref. (3):

$$S_{\mathrm{MP/GE}} = \frac{S_{\mathrm{MP}}}{S_{\mathrm{GE}}} \propto \frac{f_{\mathrm{R}}\rho\widetilde{M}_{\mathrm{z,MP}}\sin(f_{\mathrm{T}}\alpha_{\mathrm{MP}})\exp(-T_{\mathrm{E}}R_2^*)}{f_{\mathrm{R}}\rho\widetilde{M}_{\mathrm{z,GE}}\sin(f_{\mathrm{T}}\alpha_{\mathrm{GE}})\exp(-T_{\mathrm{E}}R_2^*)} = \frac{\widetilde{M}_{\mathrm{z,MP}}\sin(f_{\mathrm{T}}\alpha_{\mathrm{MP}})}{\widetilde{M}_{\mathrm{z,GE}}\sin(f_{\mathrm{T}}\alpha_{\mathrm{GE}})} \tag{1}$$

where $\rho$ denotes PD, $\widetilde{M}_{\mathrm{z,MP/GE}} = M_{z,MP/GE}/\rho$ is the longitudinal magnetization per unit PD, $\gamma$ is the gyromagnetic ratio, $f_{\mathrm{R}}$ is a factor denoting the spatial dependence of receive sensitivity (here, a weighted linear combination of individual channels). The arguments of the sine functions are the local flip angles, $\alpha_{\mathrm{loc,MP,GE}} = f_{\mathrm{T}}\alpha_{\mathrm{MP,GE}}$ where $f_{\mathrm{T}}$ denotes the transmit field ($B_1^+$) inhomogeneity. Finally, $R_2^* = 1/T_2^*$ is the effective transverse relaxation rate. Note that $\widetilde{M}_{\mathrm{z,MP}}$ is acquired under transient conditions towards a driven equilibrium, $\widetilde{M}_0^*$, with an increased rate $R_1^* = 1/T_1^*$ as (7):

$$\widetilde{M}_0^* = \widetilde{M}_0 \cdot \frac{1-\exp(-R_1 TR)}{1-\exp(-R_1^* TR)} \tag{2}$$

where $\widetilde{M}_0$ is the thermal equilibrium and $R_1^*$ is:

$$R_1^* = R_1 - \ln(\cos(f_T\alpha_{\mathrm{MP}}))/TR \tag{3}$$

and $R_1 = 1/T_1$. If full relaxation within one cycle is not obtained, $\widetilde{M}_{z,MP}$ will attain a dynamic steady state between cycles which usually occurs after a few cycles (8). How close $\widetilde{M}_{z,MP}$ will be to $\widetilde{M}_0^*$ when the central k-space line is acquired is a function of TI and $T_1$ as well as the local flip angle and thus $f_T$. On the other hand, without an inversion pulse, $\widetilde{M}_{z,GE}$ is always acquired under steady-state conditions so that $\widetilde{M}_{z,GE} = \widetilde{M}_0^*(f_T\alpha_{GE}, T_1)$. Thus, transmit field related bias is not necessarily removed for $\alpha_{MP} = \alpha_{GE}$ as Eq. (1) might imply.

## 3. Methods

The protocols were implemented on an actively shielded 7T MR system (Achieva, Philips Healthcare, Best, NL, software release R5.1.7.0 B), using a head coil with two transmit and 32 receive channels (Nova Medical, Wilmington, MA). The protocols were later transferred to release R5.1.7.0 C, but all experiments presented here were performed on the older release. Healthy adult subjects were scanned after giving informed written consent as approved by the regional Ethical Review Board. System specific procedures to speed up acquisition and reconstruction were disabled as far as possible to reduce possible interference with measured signal. These procedures included “Recon compression”, “Image filter”, “radial turbo direction”, “3D free factor” and “Elliptical k-space shutter” (quotation marks denote Philips-specific terminology). All images used the same “Uniformity” setting (“CLEAR”).

### *MPRAGE acquisition*

The MPRAGE protocol was built upon the standard protocol for structural MRI at the research site. Isotropic voxel sizes of either $0.7^3$, $0.8^3$ or $0.9^3$ $mm^3$ were acquired with a slab-selective excitation, a readout flip angle of $\alpha_{MP}$=8°, TR=8 ms, fat-water in-phase TE=1.97 ms and a water-fat-shift (WFS) of 2.0 px (503 Hz/px). For inversion, an adiabatic pulse with duration $\tau_{inv}$=22 ms and max $B_1$ amplitude of 15 μT was used. The delay from inversion to the central k-space readout (“linear profile order”) was TI=1200 ms and the time between inversions was $T_{cycle}$=3500 ms. These timings allowed for (i) a period of free relaxation after the readout train (increasing dynamic range) and (ii) ensured that the $M_z$ of CSF was close to the zero-crossing during acquisition of the center of k-space. Both (i) and (ii) will improve $T_1$ contrast on magnitude images. After each inversion, a 2D plane of k-space was acquired (“single-shot acquisition”), meaning that the inner-loop corresponded to the anterior-posterior (“turbo direction=Y”) direction with the turbo factor, TF, being identical to the acquisition matrix size in phase direction, $N_y$. Hence, switching between the different resolutions will alter TF and hence affect the $T_1$-contrast (see Results 4.3). For $(0.7\ \text{mm})^3$ resolution, a small inner-loop SENSE-factor of 1.11 had to be applied to fit the readout train within TI=1200 ms (TF=$N_y$/1.11). In the outer-loop (right-left direction), a SENSE-factor of 2.5 was applied. The FOV in the outer-loop ($FOV_{FH,AP,RL}$=230×230×180 $\text{mm}^3$) can be enlarged without affecting contrast. SENSE-related wrap-around artifacts in the AP direction was avoided through a Philips-specific oversampling margin (default setting) which did not affect acquisition time, $T_{acq}$. Lastly, to explore the possibility to further reduce $T_{acq}$, the protocol was implemented with elliptical k-space sampling in the phase-encoding directions ("elliptical k-space shutter"). A prerequisite for this kind of readout is a multi-shot acquisition as well as activation of the "3D free factor". This function allows for a "hybrid profile-order", effectively maintaining a constant TF but alternating the turbo direction between the inner and outer phase-encoding loops. This variation is performed in such a way that overall contrast is unaffected (for constant TF). The acquisition matrix, $N_{x,y,z}$, TF and $T_{acq}$ for all spatial resolutions with/without ellipsoidal k-space readout are listed in Table 2.

*Reference GE acquisition*

The reference GE was a TFE sequence without inversion pulse and recovery delay, acquired at identical TR and TE and outer-loop SENSE factor as the MP2RAGE sequence but with 50% zero-filling ("Overcontiguous slices") in the outer-loop to reduce scan time, i.e. voxel dimension in the right-left direction of 1.4, 1.6 or 1.8 mm respectively. When determining the in-plane voxel size (see Results 4.2), the "water-fat-shift" (WFS) in pixels was changed accordingly so that the absolute fat signal displacement was constant between GE and MPRAGE. Receiver gain and flip angle were calibrated for MPRAGE and then kept constant during the GE ("sameprep"). After acquisition, the volume was reconstructed to the same matrix size as the MPRAGE through zero-filling.

*Online post-processing*

To obtain normalized DICOM images, post-processing was implemented on the host computer using the ImageAlgebra tool. This allows to perform preset algebraic operations on two acquired images. First, both the MPRAGE and reference GE were reconstructed from the sagittal acquisition format to cubic 3D volumes using the VolumeView tool. This is necessary for subsequent mathematical operations. Then, using the ImageAlgebra tool, the MPRAGE volume was divided by the GE as in Eq. (1) and then multiplied by 100 to reduce discretization errors). To smooth out high frequency artifacts (resembling “ringing”), Gaussian filtering can be applied to the GE prior to division using the PicturePlus tool. Note that this could affect the cancelling of $T_2^*$ and PD contrast in the normalized image. Because the size of the kernel is fixed, the filtering must be applied repeatedly until the desired level of smoothing is obtained. The post-processing steps suggested here are listed in Table 1. All operations are performed on physical signal intensities (“floating point values”). This procedure was then stored in the examination protocol (“Examcard”) to be executed automatically after data acquisition (Figure 1). The resulting image volume is stored and exported as sagittal DICOM files where it will undergo Philips-specific scaling of the signal intensity to 12-bit integer “stored values” which needs to be reverted back to floating point if separate acquisitions are to be comparable (9).

*Table 1. Post-processing steps to perform the online calculation of the normalized MPRAGE (poor man's MP2RAGE). Left column shows the order in which the steps must be taken. Center column shows the post-processing tool to use (N/A used for the sequence item). Each process is denoted by an icon as in the ExamCard Right column shows the name of the scan item. Assuming that the initial sequences are named "1.1 MPRAGE" and "2.1 GE", the column shows the default naming of the scan items resulting from the post-processing tools. The Gaussian smoothing is applied three times in the implementation on site.*

| Order | Post-processing tool | Scan item |
| --- | --- | --- |
| 1 | N/A | 1.1 MPRAGE |
| 2 | VolumeView | 1.2 VMPRAGE |
| 3 | N/A | 2.1 GE |
| 4 | VolumeView | 2.2 VGE |
| 5 (Optional smoothing) | PicturePlus | 2.3 eVGE |
| 6 (Optional smoothing) | PicturePlus | 2.4 eeVGE |
| 7 (Optional smoothing) | PicturePlus | 2.5 eeeVGE |
| 8 | ImageAlgebra | 1.3 sVMPRAGE |

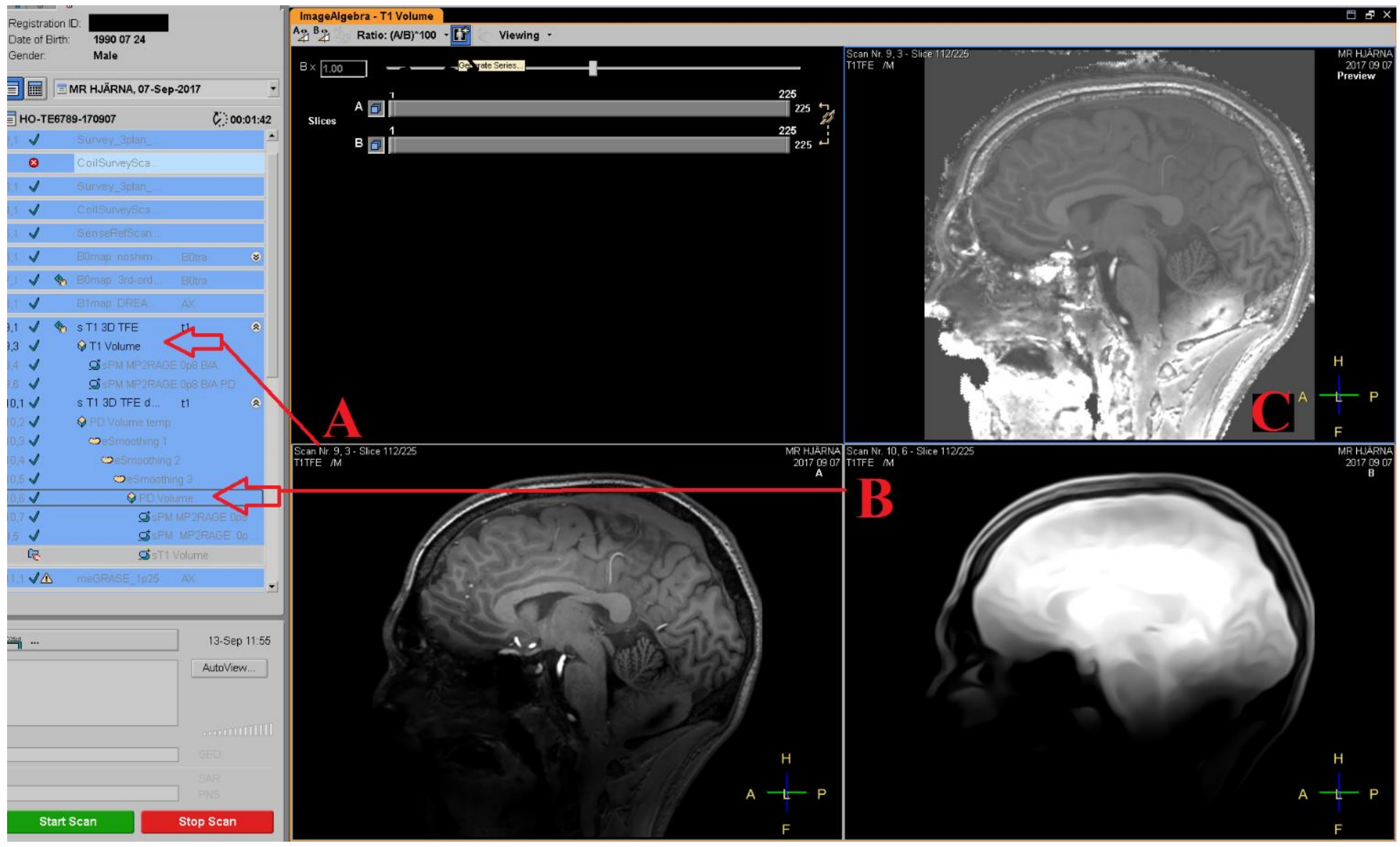

*Figure 1. Screenshot of the ImageAlgebra user interface, showing an examcard predefined to perform online post-processing to obtain a normalized MPRAGE image. An MPRAGE volume (A) and a smoothed reference GE volume (B) are connected by red lines to their respective scan item in the ExamCard (denoted by red arrows). The resulting normalized MPRAGE is shown in the upper right square (C). Note that the ImageAlgebra operation is set to "Ratio: (A/B)*100" in the upper banner.*

## *Offline post-processing*

For offline processing, DICOM images were exported, pseudo-anonymized and converted to NIfTI files using an in-house modification of the dcm2niix tool (10). The Philips-specific scaling of signal intensities was reverted from stored values to 32-bit floating point/1000 (to obtain pixel values in the range of 0-1000) and spatial dimensions were re-ordered to transverse orientation in radiological convention (right-left). Rigid co-registration of the reference GE to the MPRAGE volume (preserving the higher resolution) was performed using FLIRT (11,12) where after the normalization was performed as in Eq. (1). Segmentation of the three major tissue classes WM, GM and CSF was performed using FAST(13). Improvement in spatial homogeneity after signal normalization for different parameter settings of the reference GE, was analyzed using the coefficient-of-variation, CV, over the tissue classes in the normalized MPRAGE. The CV within a small WM ROI was used to evaluate relative changes in SNR. The rationale was that the CV in a spatially restricted ROI of homogenous tissue should be unaffected by $B_1^+$ and dominated by SNR. Average contrast between WM and GM was defined as $C = \frac{\bar{S}_{\mathrm{WM}} - \bar{S}_{\mathrm{GM}}}{\bar{S}_{\mathrm{WM}} + \bar{S}_{\mathrm{GM}}}$ where $\bar{S}_{\mathrm{WM}}$ and $\bar{S}_{\mathrm{GM}}$ is the average pixel value of the respective segmented tissue type. Further analysis of normalized MPRAGE images were performed using offline calculation rather than online post-processing.

## 3.1 Readout flip angle of reference GE

A higher $\alpha_{GE}$ will increase the $T_1$-w of the predominantly PD-w reference GE, thus reducing tissue contrast in the normalized MPRAGE image. On the other hand, reducing $\alpha_{GE}$ below the Ernst angle will decrease SNR. To find a compromise between SNR and tissue contrast in the normalized MPRAGE, $\alpha_{GE}$ was varied from 1° to 6° in increments of 1°. Reference GEs were scaled by a global factor as $S_{\mathrm{GE,scaled}}(\alpha_{\mathrm{GE,i}}) = S_{\mathrm{GE}}(\alpha_{\mathrm{GE,i}}) / \left(\alpha_{\mathrm{GE,i}} \frac{0.5 \cdot \alpha_{\mathrm{GE,6}}^2 + R_1 TR}{\alpha_{\mathrm{GE,6}}(0.5\alpha_{\mathrm{GE,i}}^2 + R_1 TR)}\right)$ to obtain comparable pixel values, where a singe $R_1 = 0.83\ s^{-1}$ was used to approximate the saturation of $M_z$. The scaling only facilitates identical windowing of the images and does not affect the analysis itself. The contrast between segmented WM and GM as well as the CV of segmented WM was plotted as a function of $\alpha_{GE}$. The latter was used as a proxy to evaluate any residual influence of $B_1^+$ inhomogeneities.

## 3.2 Voxel size of reference GRE

The $B_1$ field (both receive and transmit) is composed mostly of low spatial frequencies. Thus, to correct for $B_1$ inhomogeneity, a reference GE with low spatial resolution is sufficient. On the other hand, this is not the case for PD- and $T_2$*-contrast. However, if the normalized MPRAGE is purposed to produce only semi-quantitative images with a greatly reduced intensity field bias, some dilution of the “pure” $T_1$ contrast could be acceptable to reduce scan time. In an effort to further reduce scan time and evaluate the effect on the resulting image quality, an MPRAGE volume with 0.7 mm isotropic resolution was normalized by a reference GE where the voxel size, $V_{\mathrm{ref}}$, was varied in-plane as 0.70×0.70, 1.05×1.05, 1.40×1.40, 2.10×.2.10, and 2.80×2.80 (i.e. ×1, ×1.5, ×2, ×3, ×4 the MPRAGE resolution). The voxel dimension in the outer-loop (right-left direction) was constant at 1.40 mm resulting in acquisition times of $T_{\mathrm{acq}}$=2:21, 1:35, 1:11, 0:49, and 0:37 min, respectively

## 3.3 Implementation at different resolutions

Two finalized protocols (Results 4.1-4.2) with isotropic voxel sizes of $(0.7\ \mathrm{mm})^3$ and $(0.9\ \mathrm{mm})^3$ were used on a single subject to compare the effect of the corresponding TF on the contrast. Further, the potential increase of interpolation artifacts at different spatial resolutions (especially in the reference GE) was of interest.

## 3.4 Reproducibility

One subject was scanned on five separate occasions over a period of about 7 months using the normalized MPRAGE protocol with 0.7 mm isotropic resolution. The average CV in WM, GM and CSF before and after normalization was calculated.

## 3.5 $T_1$ calculation

As proof-of-principle, $T_1$-mapping using a LUT-based approach was performed on a healthy subject using the 0.8 mm isotropic resolution protocol together with a DREAM flip angle map. First, the evolution of the longitudinal magnetization, $M_z$, was simulated for the MPRAGE sequence with imaging parameters as described above, i.e. with $\alpha_{MP}$=8°, TR=8 ms, TI=1200 ms and TF=288.

The evolution of $M_z$ during $T_{cycle}$ in the steady state (occurring after 2-3 cycles) was simulated using Eqs. (2)-(3) in the intervals where readout occurred and normal $T_1$ relaxation where readout did not occur. Due to normalization, the simulations can be performed with $\widetilde{M}_z$ (per unit $\rho$) or $M_0$ may be applied to every pixel. The inversion efficiency applied at the end of each cycle was assumed to be $f_{inv}$=0.96 as in (4). The LUT-derived signal was then calculated for a constant $\alpha_{\mathrm{MP}}$ but a range of $f_{\mathrm{T}}$ as:

$$S_{\mathrm{MP,LUT}} = \frac{\widetilde{M}_{z,\mathrm{MP}}(f_{\mathrm{T}}, T_1, TI)}{\widetilde{M}_0} \sin(f_{\mathrm{T}} \alpha_{\mathrm{MP}}). \tag{4}$$

The reference GE was assumed to be in the steady state. Hence, $\widetilde{M}_{z,\mathrm{GE}}$ and consequently the LUT-derived GE signal, $S_{\mathrm{GE,LUT}}$, is constant for a constant $T_1$ and $f_{\mathrm{T}}$:

$$S_{\mathrm{GE,LUT}} = \frac{\widetilde{M}_{z,\mathrm{GE}}(f_{\mathrm{T}}, T_1)}{\widetilde{M}_0} \sin(f_{\mathrm{T}} \alpha_{\mathrm{MP}}). \tag{5}$$

Thus, 2D ($n_{T_1} \times n_{f_{\mathrm{T}}}$) LUTs of $S_{\mathrm{MP,LUT}}$ and $S_{\mathrm{GE,LUT}}$ are obtained for a range of $1 \leq T_1 \leq 5000$ ms (step size of 1 ms, $n_{T_1} = 5000$) and $0.4 \leq f_{\mathrm{T}} \leq 1.6$ (step size of 0.01, $n_{f_{\mathrm{T}}} = 121$). The LUTs of the two simulated signals were then combined as in Eq. (1) of Ref (4):

$$S_{\mathrm{MP2RAGE,LUT}}(T_1, f_T) = \frac{S_{\mathrm{MP,LUT}} \cdot S_{\mathrm{GE,LUT}}}{S^2_{\mathrm{MP,LUT}} + S^2_{\mathrm{GE,LUT}}}. \quad (6)$$

Thus, the values are limited to $-0.5 \leq S_{\mathrm{MP2RAGE,LUT}}(T_1, f_T) \leq 0.5$ in the final LUT for comparison with $S_{\mathrm{MP2RAGE}} = \frac{S_{\mathrm{MP}} \cdot S_{\mathrm{GE}}}{S^2_{\mathrm{MP}} + S^2_{\mathrm{GE}}}$ calculated from the measured magnitude signals. Note that the measured $S_{\mathrm{MP2RAGE}}(T_1, f_T)$ is limited to positive values ($0 \leq S_{\mathrm{MP2RAGE}} \leq 0.5$) for this implementation since it is not possible to relate the phases of two signals measured in a non-interleaved manner. The $T_1$ for which $|S_{\mathrm{MP2RAGE}} - S_{\mathrm{MP2RAGE,LUT}}|$ is minimal was calculated pixelwise either for $f_T = 1$ or as determined by the DREAM flip angle map.

# 4. Results

## 4.1 Readout flip angle of reference GE

As expected, there is a continuous decrease of tissue contrast in the normalized volume as $\alpha_{GE}$ increases (Figure 1). This decrease is evident both from visual inspection (for $\alpha_{GE}$>3°) and from the quantitative comparison of segmented WM and GM (panel C, Figure 2). The contrast was still increased after normalization compared to before for all values of $\alpha_{GE}$. The ROI analysis (panel D, Figure 2) showed no decrease in CV either at lower $\alpha_{GE}$ or after normalization, hence no apparent change in SNR could be identified. However, a minimum in the CV of segmented WM (panel E, Figure 2) is found at $\alpha_{GE}$=3°, implying minimum influence of residual $B_1^+$ inhomogeneities at this setting. The $B_1^+$ influence is visually identifiable as elevated pixel values in the center of the brain using $\alpha_{GE}$=6° (panel B, Figure 2). Based on these results, and to avoid straying too far away from the Ernst angle (~7°-5° for 1000 ms≤$T_1$≤2000 ms), $\alpha_{GE}$=3° was deemed optimal and chosen for the final protocol.

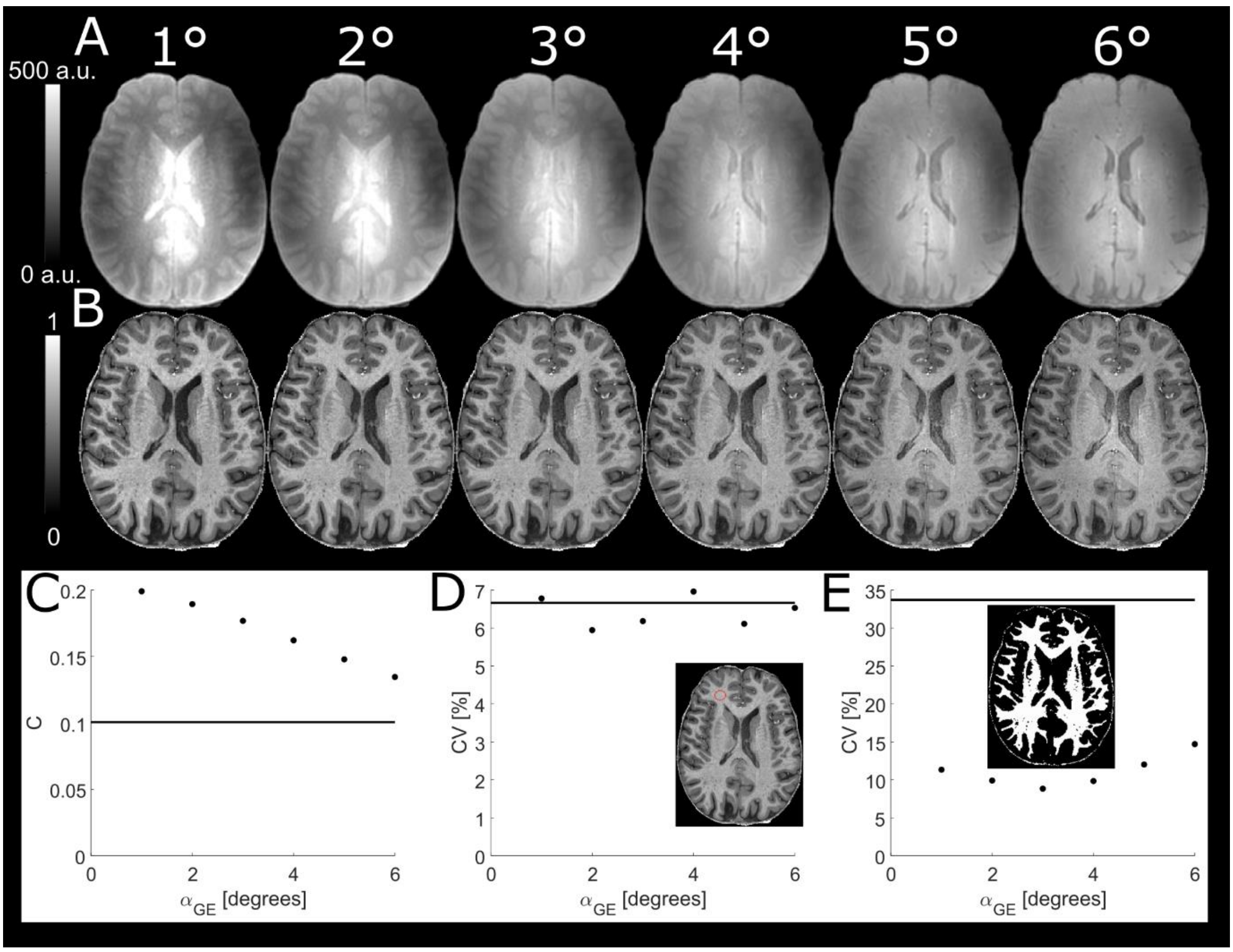

*Figure 2. Reference GEs with different $\alpha_{GE}$ (top row, A) used to obtain normalized MPRAGE volumes (center row, B). Decreasing WM-GM contrast, C, at higher $\alpha_{GE}$, is evident from visual inspection of the normalized volumes and verified in the scatter plot (C). No change in SNR could be identified by the CV in a WM ROI (red circle) (D). The CV in whole segmented WM had a minimum at $\alpha_{GE}$=3°, implying minimum influence from residual $B_1^+$ effects (E). These effects are visually identifiable as elevated pixel values in the center of the normalized MPRAGE for $\alpha_{GE}$=6°. Solid line in the scatter plots denote the MPRAGE volume before normalization.*

## 4.2 Voxel size of reference GE

Increasing $V_{ref}$ yielded very similar normalized MPRAGE images (panel B, Figure 3). Although some ringing artifacts were visible in the images (red arrow in panel A, Figure 3), this did not have a noticeably stronger effect on the normalized MPRAGE with larger $V_{ref}$. Only a slight decrease in WM-GM contrast ($C$=0.24 at $V_{ref}$=0.7×0.7×1.4 vs. $C$=0.23 at $V_{ref}$=2.8×2.8×1.4) that could possibly be due to partial volume effects (PVEs) was observable (panel C, Figure 3). Altough a higher SNR is to be expected for larger $V_{ref}$, no pattern in the CV of the WM ROI could be discerned (panel C, Figure 3). A somewhat stronger (albeit still weak) increasing trend of segmented WM CV was observed (CV=7.3% at $V_{ref}$=0.7×0.7×1.4 vs. CV=9.5% at $V_{ref}$=2.8×2.8×1.4), possibly also reflecting PVEs (panel E, Figure 3). No difference between $V_{ref}$=0.7×0.7×1.4 and $V_{ref}$=1.4×1.4×1.4 was discernible. Thus, $V_{ref}$ in one dimension was set to twice that of the MPRAGE voxel size in one dimension, i.e.

$1.4^3/1.6^3/1.8^3$ mm$^3$ for $0.7^3/0.8^3/0.9^3$ mm$^3$ respectively. A higher $V_{ref}$ was not employed to avoid stronger PVEs and interpolation errors.

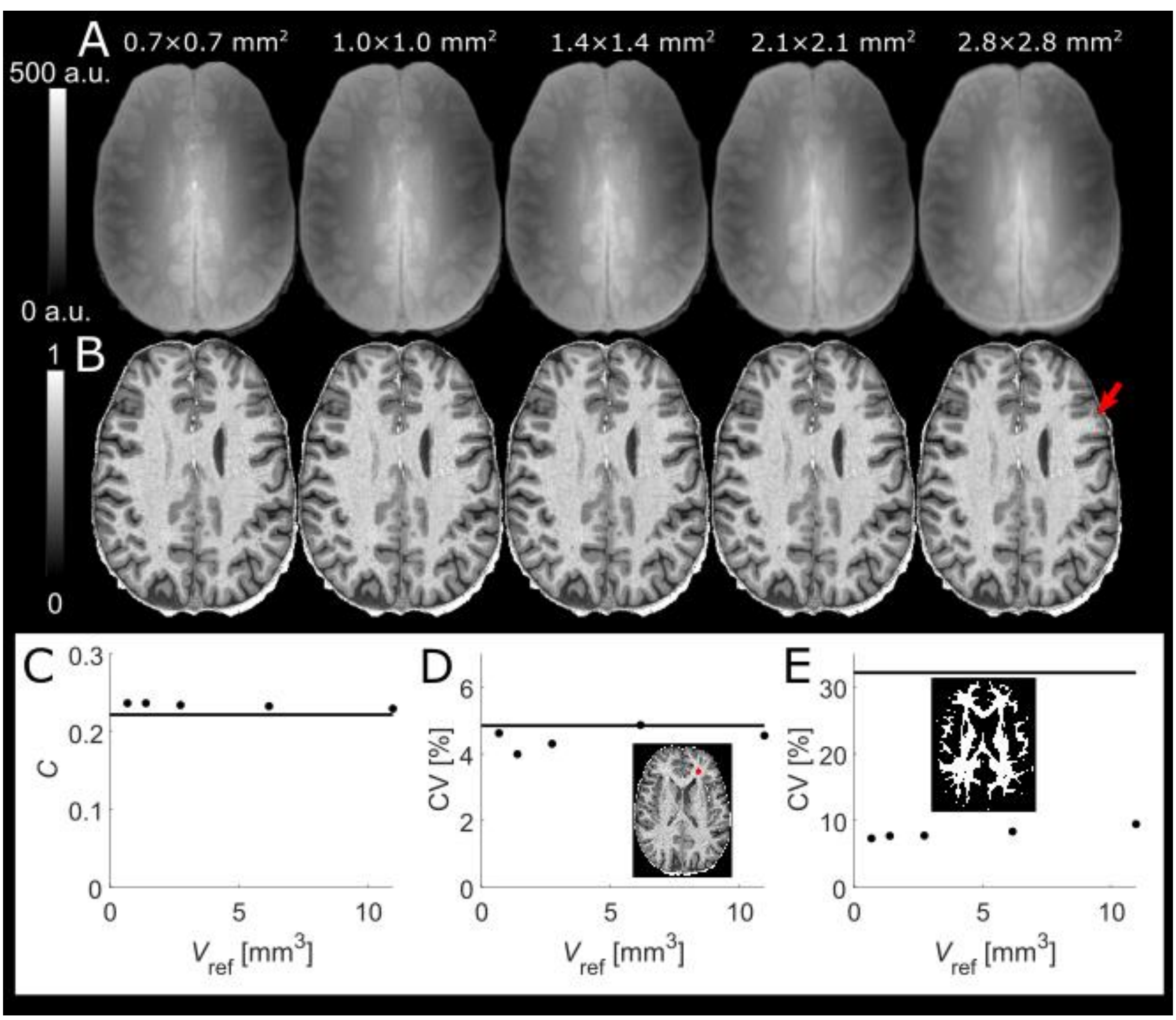

*Figure 3. Reference GEs with different in-plane voxel sizes (top row, A) used to obtain normalized MPRAGE center row, (B). The reference GEs and hence normalized MPRAGE volumes are extremely similar. Some ringing artifacts are visible (red arrow) but does not appear to severely affect image quality even at the lowest resolution in this experiment. Scatter plot of contrast vs. voxel volume (C) reveals a very weak decrease in WM-GM contrast with increasing $V_{ref}$. No change in SNR could be identified by the CV in a WM ROI (red circle) (D). A slight increase of the CV in whole segmented WM is visible at higher $V_{ref}$, possibly reflecting PVEs (E). Solid line in the scatter plots denote the MPRAGE volume before normalization.*

## 4.3 Implementation at different resolutions

Figure 4 shows normalized MPRAGE images acquired on the same subject for two of the three implemented resolutions/protocols (0.7 and 0.8 mm isotropic resolution without elliptical sampling). The signal intensity in GM remains relatively unchanged while WM becomes brighter when TF is reduced, hence increasing WM-GM contrast. The reduction in contrast due to a longer RAGE readout (higher TF) is in concordance with findings by Deichmann et al (7). Any increase in Gibbs ringing at the lower spatial resolution was not observed.

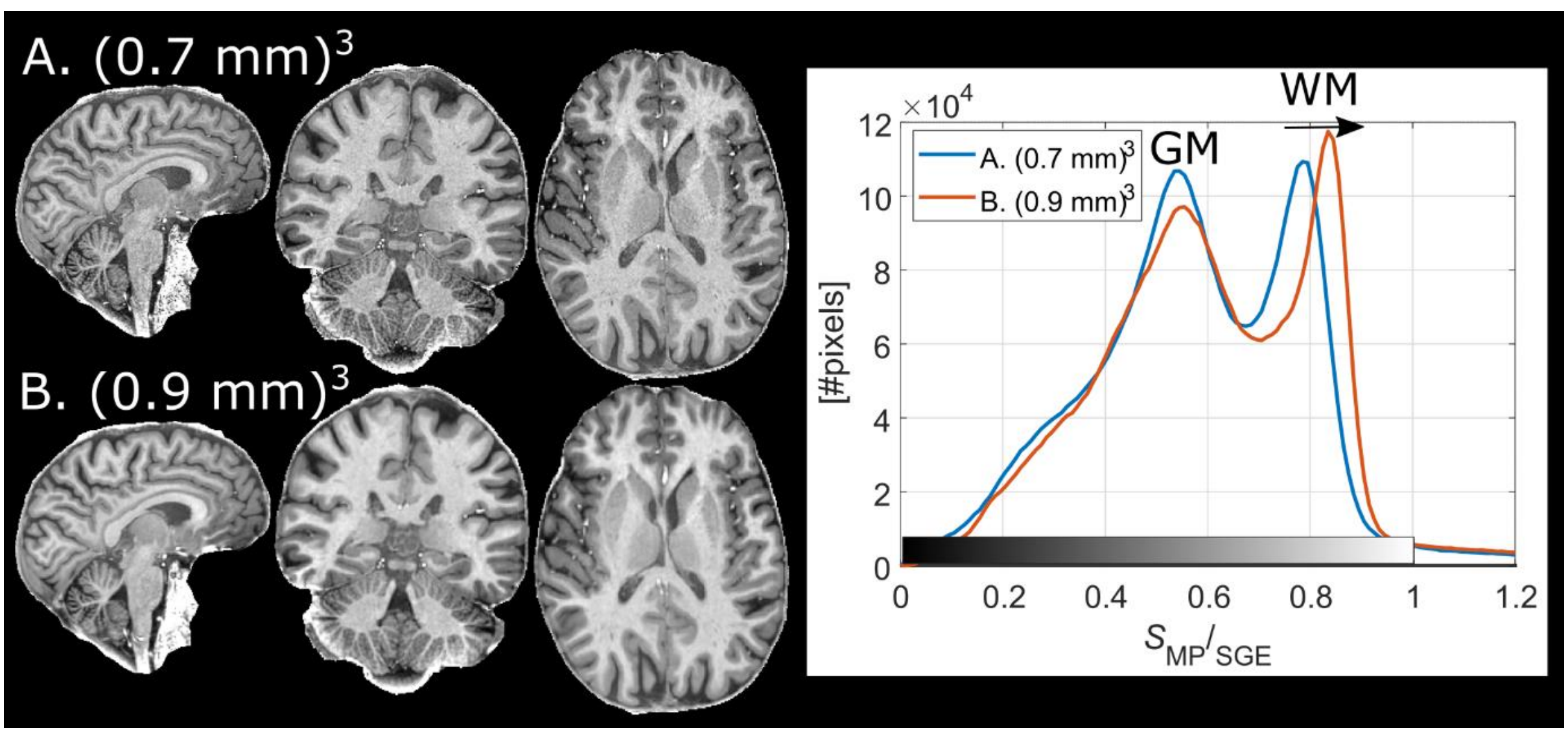

*Figure 4. Normalized MPRAGE images acquired on the same subject with two different protocols and resolutions. A. (0.7 mm)$^3$, and B. (0.9 mm)$^3$. At 0.9 mm isotropic resolution, WM pixel values are increased (black arrow in histogram plot) relative (0.8 mm)$^3$ while GM pixel values remain largely unaffected, increasing tissue contrast. This is an effect of the shortened readout train (lower TF).*

*Table 2. Acquisition time, $T_{acq}$, of the poor man's MP2RAGE protocol at different spatial resolutions and with/without free factor and elliptical k-space sampling. The MPRAGE with (0.7 mm)$^3$ resolution used an inner-loop $SENSE_{AP}$=1.11 to accommodate the readout train. The "default" option for the elliptical k-space shutter had no effect on the $T_{acq}$ of the reference GE.*

| Parameter/Resolution | (0.7 mm)$^3$ | (0.8 mm)$^3$ | (0.9 mm)$^3$ |
| --- | --- | --- | --- |
| MP $N_{x,y,z}$ | 328×328×257 | 288×288×225 | 256×256×200 |
| GE $N_{x,y,z}$ | 164×164×257 | 144×144×225 | 128×128×200 |
| MP TF | 296 | 288 | 256 |
| MP $T_{acq}$ [min] | 05:59 | 05:14 | 04:39 |
| MP $T_{acq}$ (elliptical) [min] | 04:35 | 04:04 | 03:36 |
| GE $T_{acq}$ [min] | 01:10 | 00:53 | 00:42 |
| GE $T_{acq}$ (elliptical) [min] | 01:10 | 00:53 | 00:42 |
| Total $T_{acq}$ [min] | 07:09 | 06:07 | 05:21 |
| Total $T_{acq}$ (elliptical) [min] | 05:45 | 04:57 | 04:18 |

## 4.4 Reproducibility

The normalized MPRAGE volumes acquired at each scanning session are shown in Figure 5. Before normalization, the average CVs in WM were 24±12%, 30±14% in GM and 34±16% in CSF (Figure 6). Corresponding values after normalization were 9±5% in WM, 14±8% in GM and 21±13% in CSF, implying a substantially improved reproducibility. Arguably, the reproducibility is not excellent even after normalization although it should be noted that approximately six months passed between session #1 and session #5. The residual spatial bias visible at session #4 is believed to be caused by a transmitter hardware failure with elevated pixel values representing very low values in the underlying reference GE. Bright pixel values in areas of low $B_1^+$ in the temporal lobes and cerebellum are indicative of a failed inversion with the adiabatic pulse (14). Inferior brain regions are also more susceptible to physiological noise (15). Exclusion of session #4 yielded an average CV of 20±13/26±17/28±18% in WM/GM/CSF before normalization and 8±5/11±7/18±10% after.

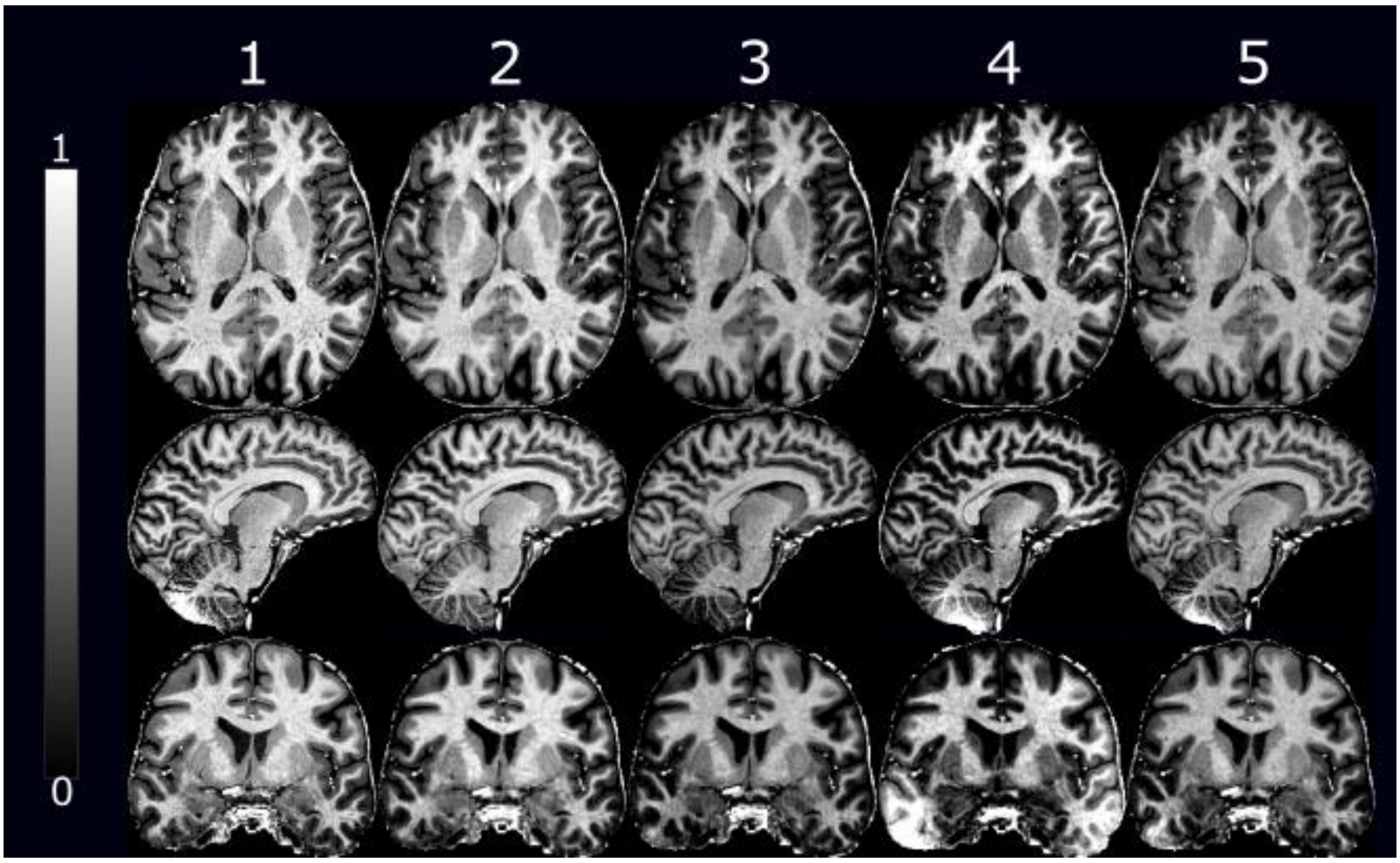

*Figure 5. Normalized MPRAGE volumes of a subject acquired at five separate scanning sessions. Reproducibility is overall high apart from low $B_1^+$ areas such as the cerebellum and temporal lobes. Bright artefacts point to lower inversion efficiency, At the $4^{th}$ scanning session, there was likely a transmitter hardware failure, resulting in generally lower $B_1^+$.*

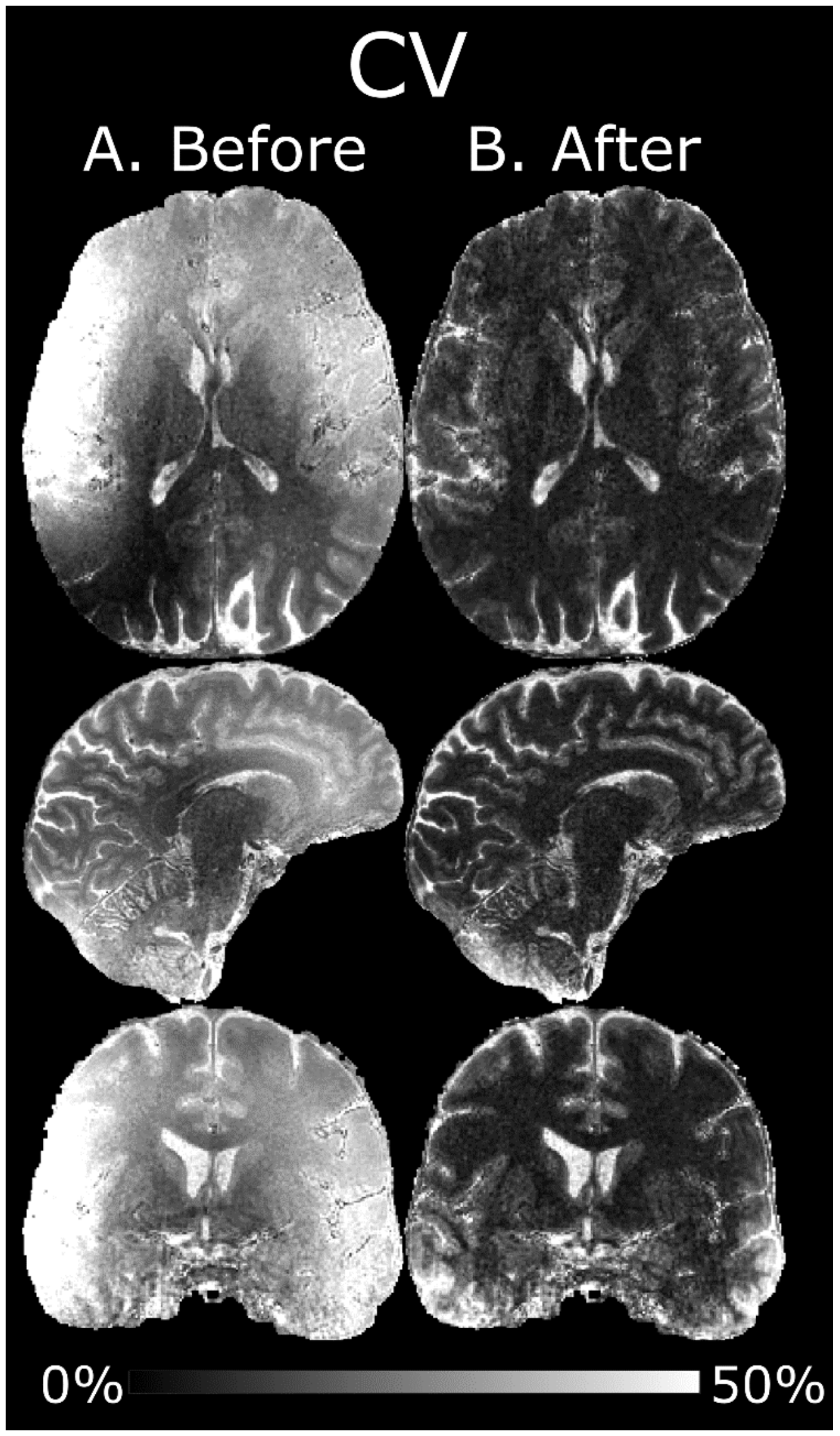

*Figure 6. The CV of the MPRAGE volumes acquired at the five scanning sessions before (A) and after (B) normalization. Reproducibility is substantially improved after normalization. The CV is noticeably higher in low $B_1^+$ regions. In addition, inferior regions are likely affected by physiological noise.*

## 4.5 $T_1$ calculation

The evolution of $\widetilde{M}_{\mathrm{z,MP}}$ over $T_{\mathrm{cycle}}$ for TF=288 is shown in panel A, Figure 7. The LUT signals were derived from this evolution for different $T_1$s and $f_{\mathrm{T}}$=1 (panel B, Figure 7). The resulting $T_1$ as a function of $S_{\mathrm{MP2RAGE}}$ at different $f_{\mathrm{T}}$-values is also shown (panel C, Figure 7). For these sequence parameters, transmit field bias is strongest at long $T_1$ (i.e. CSF). Around approximately 1000 ms, however, $T_1$ is no longer uniquely defined for a given $S_{\mathrm{MP2RAGE}}$, which thus sets the lower limit of measurable $T_1$. This lower limit varies based on $f_{\mathrm{T}}$ as 1122/1156/1181/1169/1097/966/810 ms for $f_{\mathrm{T}}$=0.4/0.6/0.8/1.0/1.2/1.4/1.6 respectively. At $f_{\mathrm{T}}$=0.4, there is also an upper limit at 3330 ms. A map of $T_1$ derived from this LUT before (panel A) and after (panel B) transmit field ($B_1^+$)-correction is shown in Figure 8. Estimated tissue $T_1$ is only moderately affected by the $B_1^+$-correction although an elevation of $T_1$ in the thalamus ($f_T$≈1.3) after correction can be discerned. The most notable difference is in the CSF which is adjusted towards higher values in high $B_1^+$ areas and towards lower values in low $B_1^+$ areas. Especially after correction, the distribution of $T_1$ looks very homogenous across different tissues without any obvious $B_1^+$ bias.

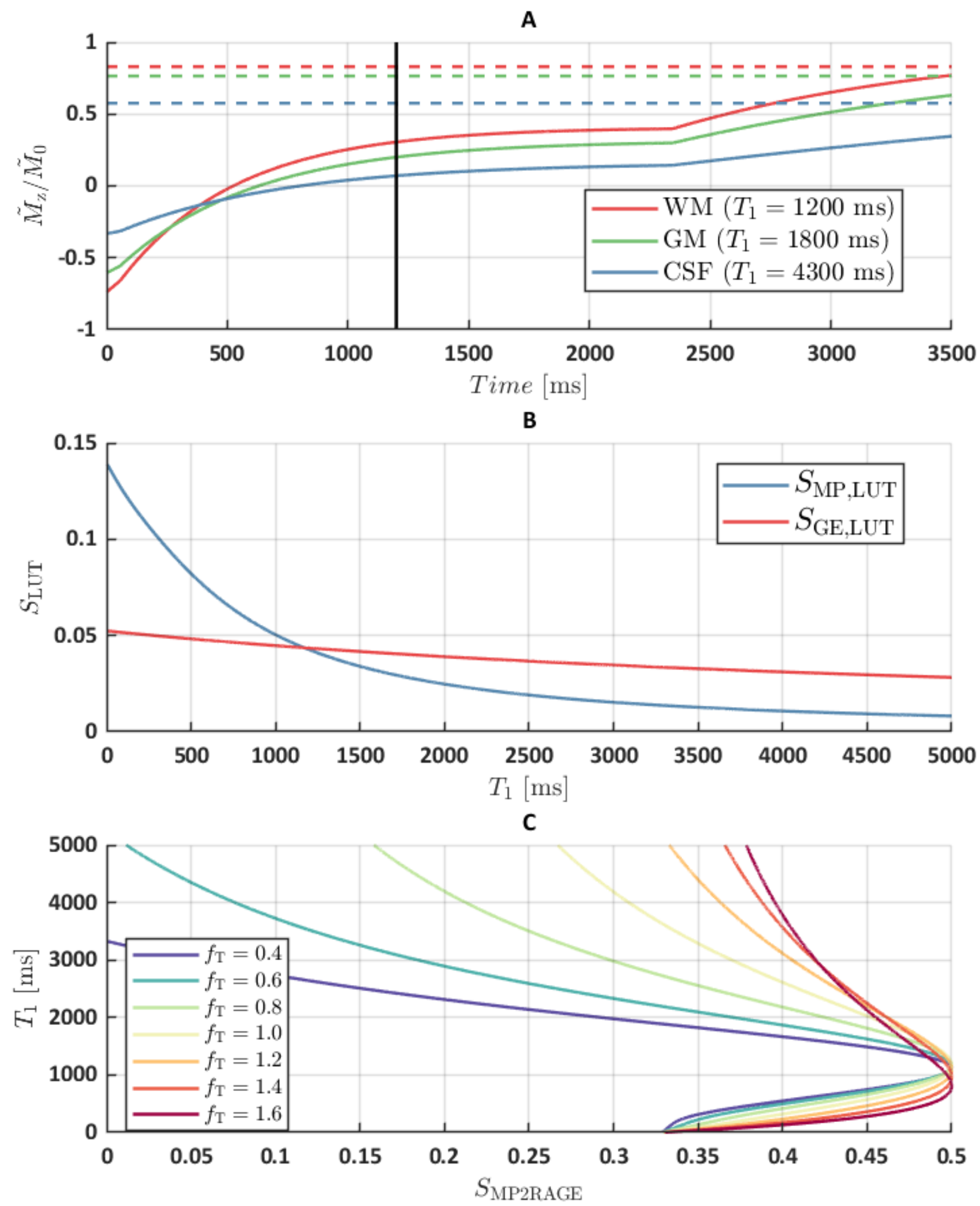

*Figure 7. A: Solid colored lines show the evolution of $M_z$ during an MPRAGE sequence across $T_{cycle}$ = 3500 ms and $f_T$=1 for three values of $T_1$ exemplifying WM, GM and CSF respectively. The dashed lines show the corresponding steady state of the reference GE. Vertical black line denotes the center of k-space at TI=1200 ms. B: The two LUT signals as a function of $T_1$ for $f_T$=1. C: Estimated $T_1$ as a function of $S_{MP2RAGE}$ for different $f_T$. Areas with longer $T_1$ are disproportionately biased by deviations in $f_T$. Depending on $f_T$, the minimum $T_1$ that can be uniquely defined range from 1181 ms ($f_T$ =0.8) and 810 ms ($f_T$ =1.6). At $f_T$ =0.4 there is also an upper limit at 3330 ms.*

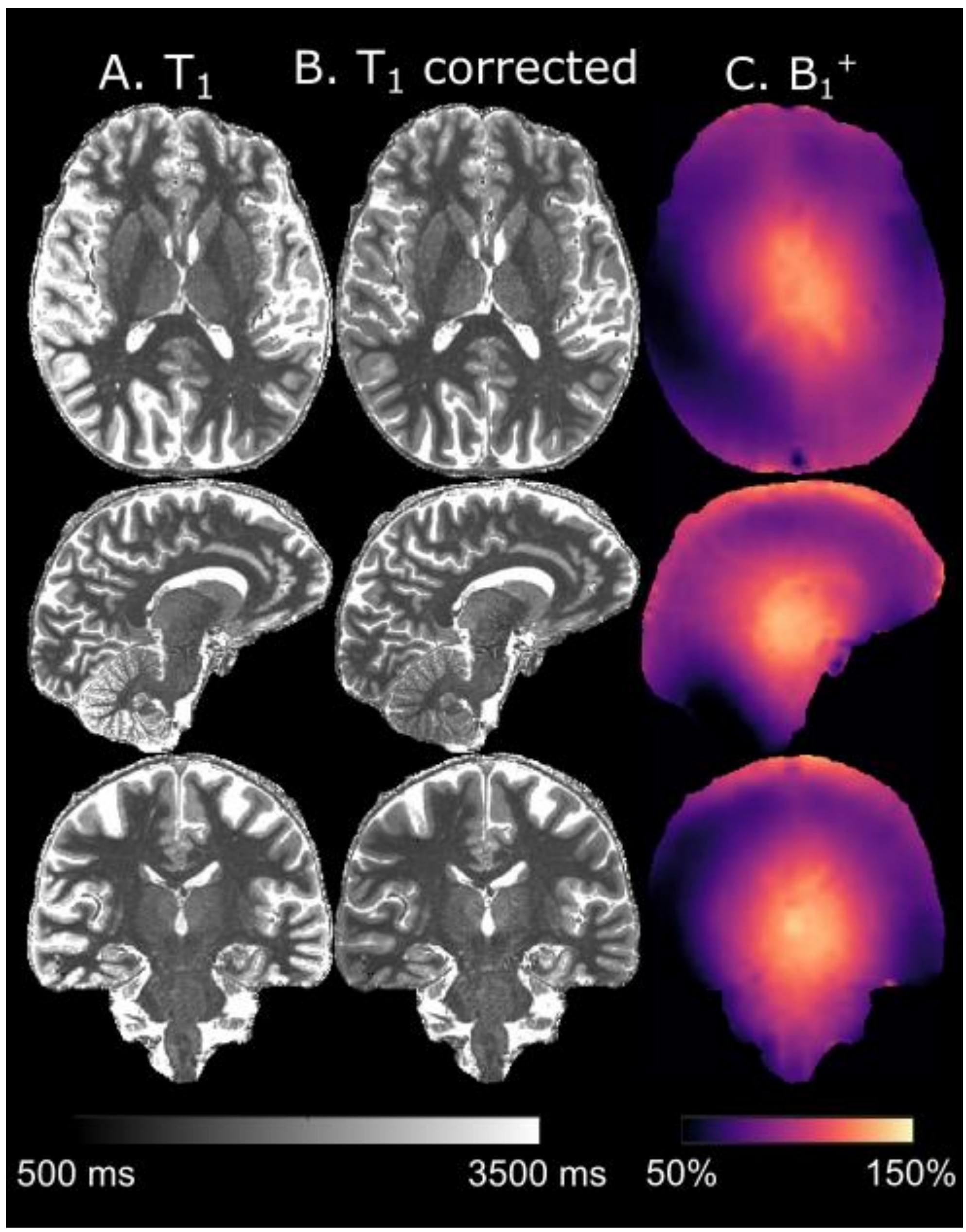

*Figure 8. LUT-derived $T_1$ maps either uncorrected (A) or corrected (B) with a separately acquired $B_1^+$ map (C). The $T_1$ estimation in tissue is moderately robust against $B_1^+$ influence and mostly it is the CSF that is affected.*

### 4.6 Example results of finalized protocol

Figure 9 shows images from a subject using the finalized protocol ($0.8^3$ $mm^3$, without elliptical sampling). After normalization, effects from the intensity field bias is considerably reduced. For instance, the diagonal $B_1^+$ pattern in the axial view is removed as well as the right-left asymmetry in the coronal view. The improved homogeneity can also be seen in the whole-brain histogram where WM and GM form distinct modes after normalization. Issues of extremely low $B_1^+$ ($f_T$~0.3) in the right part of the cerebellum where the local flip angle is too low to fulfill the adiabatic condition cannot be resolved through normalization, however.

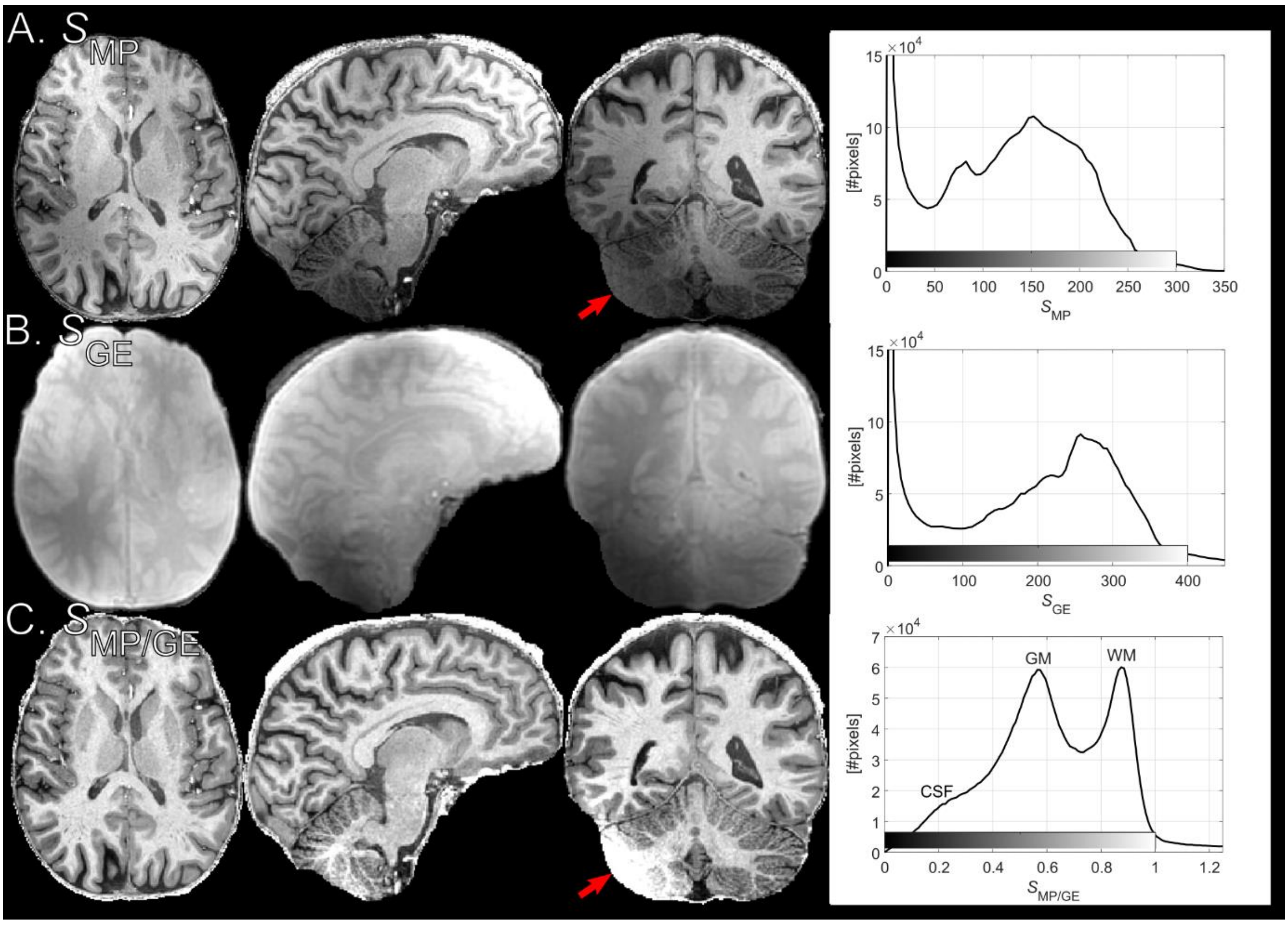

*Figure 9. Example image acquired with the MPRAGE protocol before normalization (A), the reference GE used for normalization (B) and the MPRAGE image after normalization (C). The fourth panel shows whole-brain histograms of the respective images. Normalization greatly reduces spatial heterogeneity from $B_1$. The improved homogeneity is also illustrated by the histograms where GM and WM (and to a lesser extent CSF) modes are visible after normalization. In the right part of the cerebellum (red arrow, coronal view), $B_1^+$ is very weak, leading to failed adiabatic inversion.*

## 5. Discussion

We describe the process of implementing a non-interleaved protocol for normalization of MPRAGE images at different spatial resolutions. Effects of varying the flip angle and acquisition voxel size of the reference GE was studied, mainly to improve WM-GM contrast and minimize scan time without introducing biases. The main purpose of the protocols was to facilitate obtaining semi-quantitative (i.e. reproducible) images with "pure" $T_1$ contrast that could be used for longitudinal studies or to compare acquisitions between subjects. Indeed, reproducibility on a single subject was improved after normalization, illustrated by a dramatic decrease in the tissue-specific CV (9±5% in segmented WM after normalization vs. 24±12% before).

The multiplicative spatial intensity bias imposed by the combination of receive coil signals ("Uniformity") is removed by normalization. Due to a smaller readout flip angle in the reference GE compared to the MPRAGE and to the rather small flip angles overall, the effect of the inhomogeneous transmit field ($B_1^+$) is also mostly removed. As proof of principle, we also estimated $T_1$ maps through a LUT-based approach analogous to MP2RAGE (4). As expected, accuracy was improved by using a separate $B_1^+$ map.

The MPRAGE protocol, provided by the vendor upon system installation, is comparable to that used in the Alzheimer's Disease Neuroimaging Initiative (ADNI) at 3T (16), modified to a higher spatial resolution and with prolonged sequence timings adapted for the prolonged $T_1$ at 7T. We made very little alterations to this protocol beyond what was necessary to implement different spatial resolutions. For instance, we did not try to optimize the point spread function (PSF), which will be non-ideal when signal is acquired during a transient state and can adversely affect tissue contrast (8).

The protocol was not designed for $T_1$ calculation as featured by MP2RAGE. At the rather long TI=1200 ms it fails to effectively exploit the dynamic range obtained when inverting fully relaxed longitudinal magnetization (panel A, Figure 7). This results in quite "saturated" images when normalizing using Eq. (6) (pixels close to 0.5), leading to poor tissue contrast and loss of precision in the $T_1$ calculation. Further, although whole-brain histograms of $T_1$ was very similar when using different $V_{ref}$, it is most likely prudent to use identical voxel sizes of MPRAGE and the reference GE if accurate$T_1$ estimation is of interest, especially at the cortical boundaries that are susceptible to PVEs. It is important to note that the LUT-based approach assumes that all differences in pixel values are solely due to $T_1$ (i.e. that there exists "pure" $T_1$ contrast). By design, the influence of transmit field inhomogeneities on the $T_1$ calculation is decreased by the normalization of signals. The choice of $\alpha_{MP}=2.7\alpha_{GE}$ (8° vs 3°) appeared to minimize residual effects of transmit field inhomogeneity on the normalized MPRAGE (Figure 2) and thus also on the $T_1$ calculation (Figure 8). The largest residual effect of transmit field inhomogeneity was found at $\alpha_{MP}=1.3\alpha_{GE}$ (8° vs 6°). This is in concordance with the work of Van de Moortele et al. where a choice of $\alpha_{MP}=2\alpha_{GE}$ was found to have a much stronger residual transmit field dependence than $\alpha_{MP}=\alpha_{GE}$ (3). To increase accuracy, a separately acquired flip angle map such as in ref. (17) is still recommended if $T_1$-mapping is of interest. This is especially true for longer $T_1$s where transmit field influence is stronger (Figure 7). The loss of contrast observed in the cerebellum (Figure 9) occurs when $B_1^+$ decreases below the threshold needed for the inversion pulse to fulfill the adiabatic condition and cannot be fixed by flip angle mapping (14).

With the older system software, it was not possible to conveniently implement an interleaved MP2RAGE sequence. Hence, we settled for a non-interleaved variant, aptly named the "poor man's MP2RAGE". The obvious benefit of an interleaved acquisition is that identical scanning conditions such as RF power calibration is guaranteed, as well as increased robustness against inter-scan movement. Here, inter-scan motion can be corrected by offline rigid coregistration with, for instance, FSL FLIRT (12). However, this is not possible if normalization is performed online. The reduced $T_{acq}$ of any one acquisition however will reduce the risk of intra-scan subject movement. Also, the risk of introducing $T_1$ contrast in the GE reference due to poor timings (mainly if TI and/or TF·TR is too short) is removed (3). The option to increase $V_{ref}$ also facilitates the possibility to have a shorter total scan time than needed for the interleaved MP2RAGE.

Although subdural ringing artifacts did not noticeably increase when using a protocol with lower spatial resolution (Figure 4), curved ringing artifacts were occasionally observed. These were more evident in the reference GE but could also be seen in MPRAGE (data not shown). The artifacts appeared to be correlated to subject movement, but this was not confirmed. These artifacts are believed to be related to the very low spatial resolution ($5.5\times7.4\times4.0$ mm$^3$) of the SENSE reference scan, acquired prior to the MPRAGE and reference GE. The artifact was very similar to the "streaky-linear" artifact "type A" described by Sartoretti et al. and showcased in Figure 3 panels (g), (h) and (i) (18). The artifacts could be reduced through Gaussian filtering of the reference GE as done for the online post-processing with the PicturePlus tool.

A protocol with an MPRAGE acquisition voxel size of (0.6 mm)$^3$ was also explored (data not shown). However, the SNR in the unnormalized MPRAGE was deemed unacceptably low. Hence, the (0.7 mm)$^3$ protocol here represents the upper limit on spatial resolution imposed by SNR. It should be noted that noise progression will moderately decrease the SNR in the normalized, $S_{\mathrm{MP/GE}}$, relative the unnormalized, $S_{\mathrm{MP}}$, MPRAGE by $\frac{S_{\mathrm{MP}}}{\sqrt{1+S_{\mathrm{MP/GE}}^2}}$, somewhat adversely affecting obtainable spatial resolution (3). Decreasing the acquisition voxel size also entails increasing TF and thus the duration of the readout train and eventually of TI.

However, this limitation can be circumvented by introducing a SENSE factor in the inner loop, as done here for (0.7 mm)[3]. The "3D free factor" feature of the system (allowed in a multi-shot acquisition) allows the use of a TF larger than the number of in-plane k-space lines ($N_y$), which is an effective way to decrease scan time. It further allows the enabling of the "elliptical k-space shutter", another feature which decreases $T_{acq}$ by a factor of approximately $r^2/(\pi(r/2)^2) \approx 1.3$ (Table 2). This shutter could not be employed on the reference GE by two available settings "default" and "no". When a TFE sequence is employed without an inversion pulse, the "multi-shot" option is no longer available, and the "default" setting does not activate the "elliptical k-space shutter". This could be circumvented by switching "Fast Imaging mode" from "TFE" to "none" although to avoid potentially differing reconstruction pathways for the MPRAGE and GE data, this was not implemented. This motivated us to run the MPRAGE in single-shot mode as well, however scan time can be reduced for the MPRAGE through the "elliptical k-space shutter," available for a multi-shot acquisition with the "3D free factor" turned on. On a similar note, the parameter "Overcontiguous slices" can be set to either "No" or "Yes". Thus, restricting zero-filling in the slice direction of the reference GE to a factor of 1 or 2.

## Acknowledgements

This work was supported by the Swedish Research Council (NT-2014-6193). The authors are indebted to Fredy Visser, Philips Healthcare, for providing the MPRAGE protocol and the processing routine. Lund University Bioimaging Center (LBIC) is acknowledged for experimental resources (equipment grant VR RFI 829-2010-5928).

## References

1. Mugler JP, 3rd, Brookeman JR. Three-dimensional magnetization-prepared rapid gradient-echo imaging (3D MP RAGE). Magn Reson Med 1990;15(1):152-157.
2. Balchandani P, Naidich TP. Ultra-High-Field MR Neuroimaging. AJNR Am J Neuroradiol 2015;36(7):1204-1215.
3. Van de Moortele PF, Auerbach EJ, Olman C, Yacoub E, Ugurbil K, Moeller S. T1 weighted brain images at 7 Tesla unbiased for Proton Density, T2* contrast and RF coil receive B1 sensitivity with simultaneous vessel visualization. Neuroimage 2009;46(2):432-446.
4. Marques JP, Kober T, Krueger G, van der Zwaag W, Van de Moortele PF, Gruetter R. MP2RAGE, a self bias-field corrected sequence for improved segmentation and T1-mapping at high field. Neuroimage 2010;49(2):1271-1281.

5. Focke NK, Helms G, Kaspar S, Diederich C, Toth V, Dechent P, Mohr A, Paulus W. Multi-site voxel-based morphometry--not quite there yet. Neuroimage 2011;56(3):1164-1170.
6. Trampel R, Bazin PL, Pine K, Weiskopf N. In-vivo magnetic resonance imaging (MRI) of laminae in the human cortex. Neuroimage 2019;197:707-715.
7. Deichmann R, Good CD, Josephs O, Ashburner J, Turner R. Optimization of 3-D MP-RAGE sequences for structural brain imaging. Neuroimage 2000;12(1):112-127.
8. Mugler JP, 3rd, Brookeman JR. Rapid three-dimensional T1-weighted MR imaging with the MP-RAGE sequence. J Magn Reson Imaging 1991;1(5):561-567.
9. Chenevert TL, Malyarenko DI, Newitt D, Li X, Jayatilake M, Tudorica A, Fedorov A, Kikinis R, Liu TT, Muzi M, Oborski MJ, Laymon CM, Li X, Thomas Y, Jayashree KC, Mountz JM, Kinahan PE, Rubin DL, Fennessy F, Huang W, Hylton N, Ross BD. Errors in Quantitative Image Analysis due to Platform-Dependent Image Scaling. Transl Oncol 2014;7(1):65-71.
10. Li X, Morgan PS, Ashburner J, Smith J, Rorden C. The first step for neuroimaging data analysis: DICOM to NIfTI conversion. J Neurosci Methods 2016;264:47-56.
11. Jenkinson M, Bannister P, Brady M, Smith S. Improved optimization for the robust and accurate linear registration and motion correction of brain images. Neuroimage 2002;17(2):825-841.
12. Jenkinson M, Beckmann CF, Behrens TE, Woolrich MW, Smith SM. Fsl. Neuroimage 2012;62(2):782-790.
13. Zhang Y, Brady M, Smith S. Segmentation of brain MR images through a hidden Markov random field model and the expectation-maximization algorithm. IEEE Trans Med Imaging 2001;20(1):45-57.
14. Kadhim M. Measuring T1 using MP2RAGE in Human Brain at 7T – Effect of B1+ and Inversion Pulse Efficiency2021.
15. Meineke J, Nielsen T. Data consistency-driven determination of B 0 -fluctuations in gradient-echo MRI. Magn Reson Med 2019;81(5):3046-3055.
16. Jack CR, Jr., Bernstein MA, Fox NC, Thompson P, Alexander G, Harvey D, Borowski B, Britson PJ, J LW, Ward C, Dale AM, Felmlee JP, Gunter JL, Hill DL, Killiany R, Schuff N, Fox-Bosetti S, Lin C, Studholme C, DeCarli CS, Krueger G, Ward HA, Metzger GJ, Scott KT, Mallozzi R, Blezek D, Levy J, Debbins JP, Fleisher AS, Albert M, Green R, Bartzokis G, Glover G, Mugler J, Weiner MW. The Alzheimer's Disease Neuroimaging Initiative (ADNI): MRI methods. J Magn Reson Imaging 2008;27(4):685-691.
17. Olsson H, Andersen M, Helms G. Reducing bias in DREAM flip angle mapping in human brain at 7T by multiple preparation flip angles. Magn Reson Imaging 2020;72:71-77.
18. Sartoretti T, Reischauer C, Sartoretti E, Binkert C, Najafi A, Sartoretti-Schefer S. Common artefacts encountered on images acquired with combined compressed sensing and SENSE. Insights Imaging 2018;9(6):1107-1115.